# MBE growth of cubic $Al_xIn_{1-x}N$ and $Al_xGa_yIn_{1-x-y}N$ lattice matched to GaN

**D.J. As**[*,1], **M. Schnietz**[1], **J. Schörmann**[1], **S. Potthast**[1], **J.W. Gerlach**[2], **J. Vogt**[3] and **K. Lischka**[1]

[1] University of Paderborn, Department of Physics, Warburger Str. 100, D-33095 Paderborn, Germany
[2] Leibniz-Institut für Oberflächenmodifizierung e.V., Permoserstraße 15, D-04318 Leipzig, Germany
[3] Universität Leipzig, Institut für Experimentelle Physik II, Linnéstr. 5, D-04103 Leipzig, Germany

Ternary and quaternary cubic $c$-$Al_xIn_{1-x}N$/GaN and $c$-$Al_xGa_yIn_{1-x-y}N$/GaN heterostructures lattice-matched to c-GaN on freestanding 3C-SiC substrates were grown by plasma-assisted molecular beam epitaxy. The $c$-$Al_xGa_yIn_{1-x-y}N$ alloy permits the independent control of band gap and lattice parameter. The ternary and quaternary films were grown at 620°C. Different alloy compositions were obtained by varying the Al and Ga fluxes. The alloy composition was measured by Energy Dispersive X-ray Spectroscopy (EDX) and Rutherford Backscattering Spectrometry (RBS). X-ray reciprocal space map of asymmetric (-1-13) reflex were used to measure the lattice parameters and to verify the lattice match between the alloy and the c-GaN buffer.

## 1 Introduction

The lattice mismatch of InGaN/GaN or AlGaN/GaN heterostructures introduces serious limitations for device design and/or operation. For example the large lattice mismatch in AlGaN/GaN heterostructures decreases the critical thickness of fully strained AlGaN barriers in GaN based transistors. This effect introduces relaxation via misfit dislocation and degrades the electronic transport properties [1]. To overcome these problems several groups have grown quaternary AlGaInN films by chemical vapor deposition (CVD) [1,2] or molecular beam epitaxy (MBE) [3,4,5]. These films permit the independent control of band gap and lattice parameters through changes in the quaternary composition. However, growth of quaternary AlGaInN is a challenge due to the different bond length and desorption temperature of the binary compounds. Al-containing layers normally require much higher growth temperatures than In compounds. In has a high vapor pressure and therefore the growth temperature has to be lowered in order to increase the indium incorporation and to reduce the thermal dissociation of the In-N bonds.

Recently, great interest in nonpolar III-nitrides - including cubic III-nitrides - has risen due to the absence of the built-in electrostatic field, which can limit the performance of devices. The cubic III-nitride polytype is metastable and can only be grown successfully in a narrow window of process conditions [6]. For the fabrication of electronic devices also in cubic group III nitrides it is essential to study the influence of strain and to grown lattice matched cubic AlGaInN and AlInN epilayers on GaN.

[*] Corresponding author: e-mail: d.as@uni-paderborn.de, Phone: +49 5251 60 5838, Fax: +49 5251 60 5843

In this work we demonstrate the growth of lattice matched cubic quaternary $Al_xGa_yIn_{1-x-y}N$/GaN heterostructures on freestanding 3C-SiC (001) substrates by plasma assisted molecular beam epitaxy (PAMBE).

## 2 Experimental

The samples were grown on freestanding 3C-SiC (001) substrates by molecular beam epitaxy. An *Oxford Applied Research* HD25 radio frequency plasma source was used to provide activated nitrogen atoms. Indium, aluminum and gallium were evaporated from Knudsen cells. Cubic GaN layers were deposited at 720°C directly on 3C-SiC substrates under well defined growth conditions [7]. Quaternary c-$Al_xGa_yIn_{1-x-y}N$ layers were deposited on a 500nm c-GaN buffer at 620°C in order to increase the In incorporation [8] and to reduce the thermal dissociation of In-N bonds. Different alloy compositions were obtained by varying the Ga to Al flux ratio and the In flux while keeping the substrate temperature constant.

The alloy composition was measured by Rutherford Backscattering Spectrometry (RBS) and Energy Dispersive X-ray Spectroscopy (EDX) and the lattice parameters of the layers were analyzed by High Resolution X-ray Diffraction (HRXRD).

## 3 Results and Discussion

The growth of III nitrides e.g. GaN, AlN is performed at high growth temperatures to ensure sufficient diffusion length of Ga and Al during epitaxy. Therefore the maximum growth temperature for $Al_xGa_yIn_{1-x-y}N$ is in principle determined by the incorporation of In in AlGaN, which is negligible at growth temperatures for c-GaN (720°C). For that reason it is necessary to decrease the growth temperature in order to increase the In incorporation. From previous studies for InGaN we have established that at 620°C In can be incorporated up to a molar fraction of 0.20. However at these substrate temperatures Al and Ga re-evaporation is negligible i.e., their sticking coefficient is unity. Due to the fact that the Al-N and Ga-N bonds are stronger than the In-N bond the sum of the Al and Ga fluxes remains below the active nitrogen flux, while the metal excess is provided by the In flux. For that reason we grow cubic $Al_xGa_{1-x}N$ with an Al-content of about 0.22 at a reduced temperature of 640°C in a first step. Clear RHEED oszillations (not shown here) could be observed, from which a growth rate of 163 nm/h could be estimated. However, in comparison with equivalent $Al_xGa_{1-x}N$ epilayers grown at 720°C the full width of half maximum (FWHM) of rocking curve increases from 20 arcmin to 31 arcmin demonstrating a slight degradation of crystal quality.

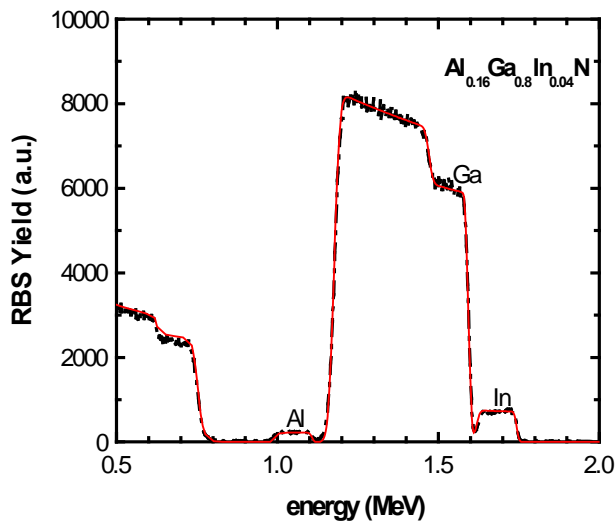

**Fig. 1** Measured and simulated (full curve) RBS spectrum of an $Al_{0.16}Ga_{0.8}In_{0.04}N$ sample grown on c-GaN/3C-SiC at substrate temperature of 620°C. The surface channels are marked by the corresponding element.

The alloy composition of the quaternary films was determined by Energy Dispersive X-ray Spectroscopy (EDX) and by Rutherford backscattering spectrometry (RBS) with a 2 MeV He$^+$ ion beam. The energy detector angle was 170° and the detector resolution 13keV. To avoid channelling, the samples were tilted 5° of the sample normal and randomly rotated. Figure 1 shows the measured and the simulated RBS spectra of an $Al_{0.16}Ga_{0.8}In_{0.04}N/GaN$ hetero-structure. The simulation fits accurately the experimental data, which demonstrates homogeneity and the precise determination of the alloy composition by RBS. The thickness of the quaternary layer estimated by RBS is 120 nm. Within experimental errors EDX measurements gave similare atomic concentrations as RBS.

With the lattice parameters of cubic nitrides $a_{AlN}$=4.38Å, $a_{GaN}$=4.52Å and $a_{InN}$=4.98Å [6], the lattice parameter of the quaternary cubic $Al_{0.16}Ga_{0.8}In_{0.04}N$ layer can be calculated by Vegard's law

$$a_{AlGaInN} = xa_{AlN} + ya_{GaN} + (1-x-y)a_{InN} \qquad (1)$$

Using the composition determined by RBS a lattice constant $a_{AlGaInN}$=4.516Å of the $Al_{0.16}Ga_{0.8}In_{0.04}N$ sample is calculated, which is nearly identical to that of the GaN buffer.

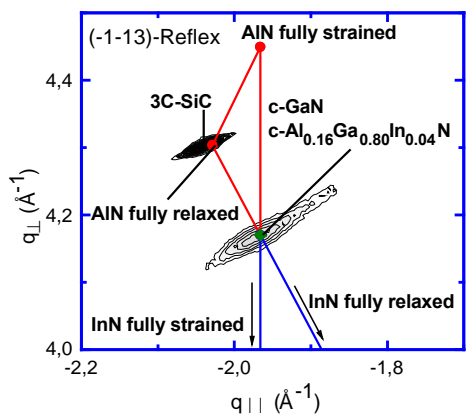

**Fig. 2** Reciprocal space map of the (-1-13)-reflex of a lattice matched quaternary cubic $Al_{0.16}Ga_{0.80}In_{0.04}N/GaN$ heterostructure.

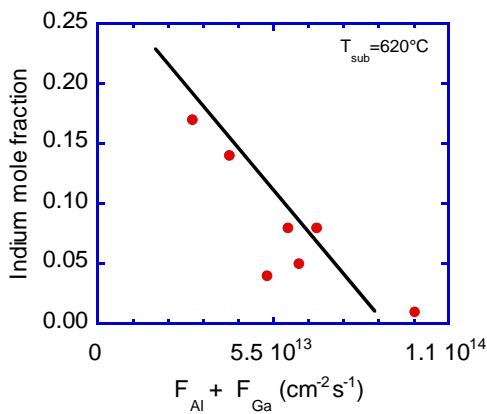

**Fig. 3** In content as a function of Al and Ga fluxes for $Al_xIn_yGa_{1-x-y}N$ films. A constant In-flux ($F_{In}$= 1.7x10$^{14}$cm$^{-2}$s$^{-1}$) and a substrate temperature of 620°C were used.

In order to determine the lattice parameter and the lattice match of the quaternary layers, reciprocal space maps of asymmetric (-1-13) reflexes were performed. Figure 2 shows the reciprocal space map (RSM) of the asymmetric (-1-13) reflex of the same sample as used in Fig 1. The red and blue triangles stretches the possible positions of the (-1-13) reflexes of various cubic quaternary $Al_xGa_yIn_{1-x-y}N/GaN$ films. The theoretical point of lattice matched cubic $Al_xGa_yIn_{1-x-y}N$ derived by equation (1) is indicated as a green point in the figure (crossing of red and blue lines). The vertical lines indicate the positions of fully strained layers whereas the inclined lines correspond to fully relaxed quaternary $Al_xGa_yIn_{1-x-y}N$ layers. In Fig. 2 two Bragg reflexes, one of the 3C-SiC substrate and one which is due to the c-$Al_{0.16}Ga_{0.80}In_{0.04}N/GaN$ heterostructure are observed. The superposition of the c-$Al_{0.16}Ga_{0.80}In_{0.04}N$ and the c-GaN Bragg peak clearly demonstrates, that both layers have identical lattice parameters. Therefore, we conclude that the quaternary layer is lattice matched to the c-GaN buffer.

For samples grown under the same In-flux and substrate temperature but with different Al to Ga flux ratio lattice matching was not fulfilled and a third reflex appeared in the reciprocal space map (not shown here). These reflexes were within the red triangle for increased Al to Ga flux ratios and shifted to the blue triangle for decreased Al to Ga flux ratios.

As mentioned before, due to the stronger bonds of Al-N and Ga-N in comparison to the the In-N bond the sum of the Al and Ga fluxes determines the maximum incorporated In in the quaternary $Al_xGa_yIn_{1-x-y}N$ epilayers. In Fig. 3 the maximum In content of several lattice matched and lattice mismatched quaternary $Al_xGa_yIn_{1-x-y}N$ epilayers grown at 620°C and with a constant In-flux of $F_{In}=1.7 \times 10^{14} cm^{-2}s^{-1}$ is plotted versus the sum of the Al and Ga fluxes. With increasing Ga and Al flux a clear predatory competition between the metal elements occur and Ga and Al atoms edges out the In atoms. For 620°C, metal rich growth conditions and the supplied In-flux the incorporated In concentration is limited to about 0.27 (extrapolated In value of the full line in Fig 3 to $F_{Al} + F_{Ga} = 0$). Similar observations have also been observed in the ternary compounds InGaN [9] and AlGaN [10].

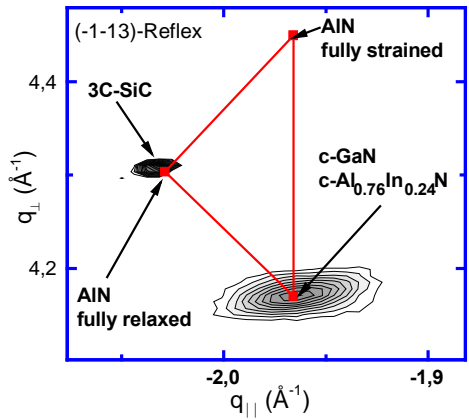

**Fig. 4** Reciprocal space map of the (-1-13)-reflex of a ternary cubic $Al_{0.76}In_{0.24}N/GaN$ heterostructure.

From Equation (1) the lattice match to c-GaN is fulfilled if for a given Al content x the Ga content is 1-1.31x and the In content is 0.31x [11]. Therefore, for the lattice matched ternary AlInN compound the In content has to be as high as 0.24 and the substrate temperature of 620°C is just sufficient to enable the incorporation of this In amount. Fig. 4 shows the RSM of such a lattice matched ternary cubic $Al_{0.76}In_{0.24}N/GaN$ heterostructure.

## 4 Conclusion

Cubic quaternary AlGaInN and ternary AlInN lattice matched to c-GaN were grown by plasma-assisted molecular beam epitaxy on free standing 3C-SiC substrates. Different quaternary alloy compositions were achieved by changing the Al to Ga flux ratio, while keeping the substrate temperature constant. The ternary and quaternary films were grown at 620°C. The alloy composition was measured by RBS. X-ray RSM of asymmetric (-1-13) reflex verify the lattice match between the alloy and the c-GaN buffer.

**Acknowledgements** We would like to thank Dr. H. Nagasawa and Dr. M. Abe from HOYA Corporation for supplying the 3C-SiC substrates.